\documentclass[conference]{IEEEtran}

\usepackage[USenglish]{babel}
\usepackage[utf8]{inputenc}
\usepackage{amstext}
\usepackage{amsmath}
\usepackage{graphicx}
\usepackage{amsfonts}
\usepackage{amssymb}
\usepackage{enumerate}
\usepackage{mathdots}
\usepackage{comment}
\usepackage{setspace}
\usepackage{float}
\usepackage{multicol}
\usepackage{slashbox}
\usepackage{multirow}
\usepackage{hhline}
\usepackage{subfigure}

\newcommand{\brackc}[2]{\lfloor #1\rceil^{#2}}

\newcommand{\RE}{\mathbb{R}}

\newtheorem{remark}{Remark}

\begin{document}
\title{Chattering analysis of Lipschitz continuous sliding-mode controllers}

\author{\IEEEauthorblockN{C. Arturo Mart\'{i}nez-Fuentes\IEEEauthorrefmark{1},
		Ulises P\'{e}rez Ventura\IEEEauthorrefmark{2}, Leonid Fridman\IEEEauthorrefmark{3}}
	\IEEEauthorblockA{Posgrado de Ingenier\'{i}a.
	Universidad Nacional Aut\'{o}noma de M\'{e}xico\\
		Email: \IEEEauthorrefmark{1}c.arturo.mtz@gmail.com,
		\IEEEauthorrefmark{2}ventury.sk8@gmail.com,
		\IEEEauthorrefmark{3}lfridman@unam.mx}}

\maketitle

\begin{abstract}
Lipschitz continuous sliding-mode controllers (LCSMC) are developed as the integral of discontinuous SMC, producing control signals of finite slope. Nevertheless, LCSMC still generate chattering in the presence of fast parasitic dynamics. In this paper, an analysis of chattering in systems driven by LCSMC is performed using the Harmonic Balance (HB) approach. Two kinds of LCSMC are considered: the first one is based on a linear sliding variable (LSV) and the second one on a terminal switching variable (TSV). Predictions of the amplitude and frequency of self-excited oscillations allowed to compute the average power consumed by the controller, in order to maintain the trajectories into the real sliding mode. A comparison of LCSMC with the Super-Twisting controller (STC), which produce a continuous control signal with infinite slope, is performed. Theoretical predictions and simulation results confirm that LCSMC may induce fast-oscillations (chattering) of smaller amplitude and average power than those ones caused by the STC. But, surprisingly, the chattering generated by LSV-LCSMC could be smaller than that one caused by TSV-LCSMC, when the actuators are fast enough. On the other hand, it tuns that if the sliding dynamics of the LSV-LCSMC closed-loop is of similar speed as the actuators dynamics, the system can loose even practical stability.
\end{abstract}

\section{Introduction}\label{Sec1}
\subsection{State of art}
The sliding-mode control (SMC) has been commonly used for compensation of perturbations and uncertainties \cite{Utkin92,Shtessel14}. The first-order sliding mode controllers (FOSMC) \cite{Utkin92} are usually \textit{discontinuous} at zero of the sliding outputs of relative degree one. However, the presence of unmodeled (fast) dynamics, that modify the relative degree, induces high frequency oscillations of bounded magnitude in the vicinity of zero of the sliding outputs \cite{Shtessel96,Fridman01,Levant01,Boiko09b,Levant10}. This closed-loop behavior is called chattering.

In order to avoid the chattering, the Super-Twisting controller (STC) \cite{Levant93,Levant98}, generating a \textit{continuous} control signal, was suggested to substitute discontinuous control inputs by continuous ones. However, STC contains fractional exponents (with \textbf{infinite-slope at zero}) causing chattering. Prof. Utkin proposed an example (see \cite{Utkin16}, Section 5) in which the STC produces fast-oscillations of larger amplitude than those one caused by the FOSMC. This example motivates a Harmonic Balance (HB) analysis \cite{Ventura18,Ventura19} of chattering generated by critically damped second-order actuators parameterized by their time-constant. It turns that, for fast-actuators the chattering generated by STC has smaller amplitude that one caused by FOSMC, and for sufficiently slow-actuators the amplitude of chattering generated by FOSMC is smaller than one caused by STC. Moreover, the critical value of the actuator time-constant was found for which the amplitude of chattering predicted by both controllers is the same. 

It was natural, as a next step, to analyze SMC with \textbf{finite-slope at zero}, which may adjust the chattering better. Such controllers was introduced in \cite{Levant93,Bartolini98} and they will called the \textit{Lipschitz continuous} sliding-mode controllers (LCSMC) in this paper. LCSMC can be formed by the integral of FOSMC with a linear sliding variable (LSV) or by the integral of discontinuous second-order SMC, i.e. Twisting\cite{Levant93}, Terminal\cite{Zhihong94,Fridman15}, Sub-optimal\cite{Bartolini98} and Quasi-continuous\cite{Levant05b} controllers. The disadvantage of such controllers is that sliding variables contains information of the output and its time-derivative. If a linear sliding variable (LSV) \cite{Utkin92,Shtessel14} is selected, the output is driven to zero exponentially into sliding-modes \cite{Shtessel14}. But if a second-order SMC is chosen the finite-time convergence of the output is ensured \cite{Zhihong94}. To compare finite-time convergent LCSMC with exponentially convergent ones, Prof. Utkin chose the Terminal switching variable (TSV) (see \cite{Utkin16}, Section 4). He proposed an example showing that LSV-LCSMC produce fast-oscillations of larger amplitude than those one caused by TSV-LCSMC. 

The goal of this paper is to analyze the chattering parameters produced by LSV-LCSMC and TSV-LCSMC in such systems with fast-parasitic dynamics, using the Harmonic Balance (HB) approach.

\subsection{Methodology}
The steady-state response of the closed-loop system with SMC is composed by self-excited oscillations caused by the imperfections in the switching terms. The Describing Function (DF) approach has been commonly used to linearize the SMC constrained to a sinusoidal response. Then, the HB method can be used to predict the parameters of fast-oscillations caused by parasitic dynamics \cite{Utkin16,Ventura18,Ventura19,Boiko07}.

In this paper a DF-HB approach is used to predict the chattering parameters produced by LSV-LCSMC and TSV-LCSMC in systems with fast-parasitic dynamics. 

\subsection{Contribution}
In this paper, the chattering parameters generated by the LCSMC with LSV and TSV are analyzed using the HB approach. With this aim:
\begin{itemize}
	\item[1.] The DF's of the LCSMC with LSV and TSV, are computed for the first-time, in order to predict the amplitude and frequency of self-exited oscillations caused by the presence of fast-parasitic dynamics.
	\item[2.] The average power \cite{Ventura19} consumed by the controller to maintain the trajectories into the real sliding-mode is calculated with the knowledge of amplitude and frequency parameters.
	\item[3.] A comparison of the LCSMC with the STC in terms of chattering parameters is provided using both: HB analysis and simulations. 
\end{itemize} 
Performed analysis confirms the hypothesis formulated by Prof. Utkin in  Section 4 of \cite{Utkin16}, that the finite-time convergence is not a critical issue for LCSMC. The amplitude of chattering generated by LSV-LCSMC is smaller than one caused by TSV-LCSMC, at least for fast-actuators. 
.
\subsection{Structure of the paper}
Section \ref{Sec2} presents the problem statement, Section \ref{Sec3} summarizes the Harmonic Balance tools required to make the comparisons. The DF computation for the LSV-LCSMC and the TSV-LCSMC is in Section \ref{Sec4}. Main results are presented in Section \ref{Sec5} and simulations comparing the STC and the LCSMC are in Section \ref{Sec6}. The conclusions are listed in Section \ref{Sec7}.

\section{Problem Statement}\label{Sec2}
\subsection{Ideal Sliding-Modes}
Consider a SISO system of relative degree one with respect to the output $x(t)$ and the control input $u(t)$, 
\begin{equation}\label{System}
\dot{x}(t)  = f(t) + u(t) 
\end{equation}
where $f(t)$ is a perturbation assumed to be Lipschitz continuous, i.e. $|\dot{f}(t)|\leq\Delta$ with upperbound $\Delta$ known. The measurements of $x(t)$ and its time-derivative are available\footnote{The knowledge of $\dot{x}$ does not imply the direct compensation of the perturbation $f(t)$ by the control input $u(t)$, i.e. $u(t)=\dot{x}(t)-f(t)$ $\Rightarrow$ $\dot{x}(t)=\dot{x}(t)$.} for all $t>0$. Also, the notation $\brackc{\cdot}{0}=|\cdot|^p sign(\cdot)$ is used. As suggested in\cite{Utkin16}, it is possible to select any of the following sliding outputs:
\begin{itemize}
	\item \textbf{Linear sliding variable (LSV)},
	\begin{equation}\label{Sigma1}
	\sigma(t)  = \dot{x}(t) + b\, x(t) \,, \hspace{2mm} b>0 \,.
	\end{equation}
	\item \textbf{Terminal switching variable (TSV)},
	\begin{equation}\label{Sigma2}
	\sigma(t)  = \brackc{\dot{x}(t)}{2}+ bx(t)   \,, \hspace{2mm} b>0 \,.
	\end{equation}
\end{itemize}
For readability sake, in the later the variable $t$ will be omitted. The Lipschitz continuous sliding-mode controller (LCSMC), based on the integral of \textit{sign} function, can be expressed as
\begin{equation}\label{Control}
\dot{u} = -k \,\brackc{\sigma}{0}\,,
\end{equation}
with $k>0$ the gain of the controller. Define the variables $x_1=x$ and $x_2 = f + u$, then, it is possible to rewrite the closed-loop system as 
\begin{equation}\label{CL}
\begin{array}{rcl}
\dot{x}_1 & = & x_2 \,, \\	
\dot{x}_2 & = & - k \,\brackc{\sigma}{0}  + \dot{f} \,.
\end{array} 
\end{equation}
To reject perturbation $\dot{f}$ in the second channel of (\ref{CL}), the controllers gains must be selected as $k>\Delta$. LSV-LCSMC ensure exponential stabilization of the output $x$ with relative degree one\cite{Shtessel14}. On the other hand, TSV-LCSMC for any $b$ ensure the finite-time convergence of the $x$ \cite{Zhihong94,Shtessel14}.

\begin{figure}[t]
	\begin{center}
		\vspace{0mm}
		\includegraphics[scale=0.24]{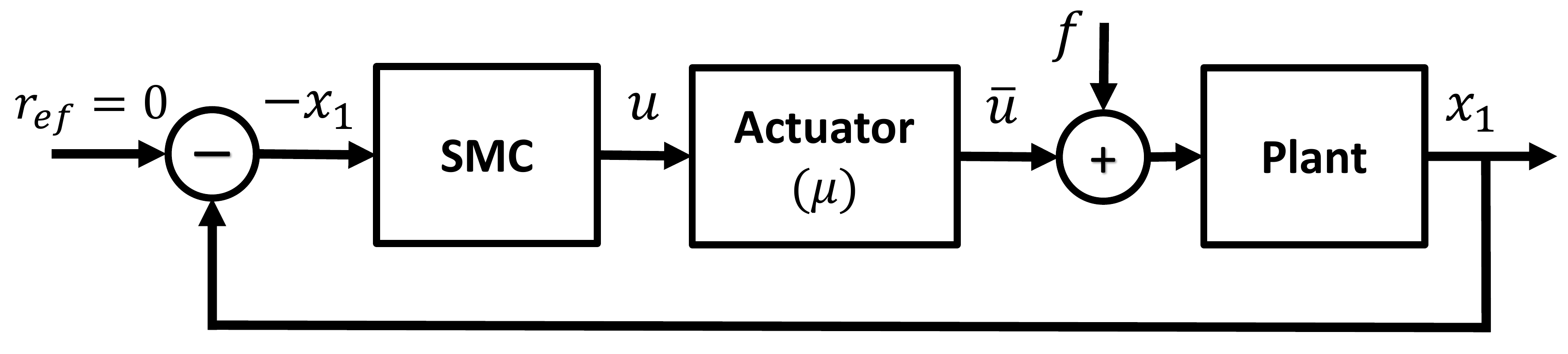}   
	\end{center}
	\vspace{-2mm}
	\caption{The SMC signal is transferred to the plant through the actuator dynamics.} \label{Bloques}
	\vspace{-2mm}
\end{figure}

\subsection{Real Sliding-Modes}
The presence of unmodeled (parasitic) dynamics is inevitable in real applications such that relative degree could be uncertain. Fast-parasitic dynamics could be excited by the infinite gain of the SMC laws, so that chattering (undesired oscillations in a vicinity of the sliding surface $\sigma=0$) appears\cite{Levant10}, even with continuous SMC \cite{Ventura19,ByF05}. To study the effect of parasitic dynamics on the control loop, a system like in Figure \ref{Bloques} will be considered, where the effect of the input (\ref{Control}) to the plant (\ref{System}) is affected by an actuator. The cascade connection of the actuator and the plant can be rewritten as 
\begin{equation}\label{ActuatorModel}
\begin{array}{rcl}
\dot{x}_1 &=& x_2 \,, \\
\dot{x}_2 &=& \dot{f} + \bar{u} \,, \\
\mu\dot{z} &=& g(z,u) \,, \hspace{3mm} \bar{u} = h(z) \,, 
\end{array}
\end{equation}
where $z\,\in\,\RE^m$ is the actuator state, $\bar{u}\,\in\,\RE$ is the output of the actuator and $u\,\in\,\RE$ is the input based on sliding-mode control (\ref{Control}). The actuator dynamics assumed to be stable and such that, for small values of the actuator's time constant $\mu>0$, the output $\bar{u}$ uniformly tends to $u$. Any \textit{stable} transfer function $G_a(s)$ such that $G_a(0)=1$ can be taken as a linearized model of actuator dynamics in (\ref{ActuatorModel})\cite{Levant10}. \\

Real sliding-modes \cite{Shtessel14,Levant93} in the system contains two components of oscillations: fast-motions are the self-excited oscillations caused by the actuator fast-dynamics. Slow-motions are the propagation of external inputs since the controllers are not able to reject them exactly \cite{Boiko09b}. However, the study of fast-oscillations caused by fast-parasitic dynamics is the aim of this paper so that $f(t)=0$ is considered.

In this paper a frequency domain analysis is proposed to estimate the chattering parameters of the self-excited oscillations caused by the LCSMC (\ref{Control}) in the presence of fast-parasitic dynamics (actuator) in the system (\ref{ActuatorModel}), considering the LSV (\ref{Sigma1}) and the TSV (\ref{Sigma2}), respectively. The estimation of amplitude and frequency of self-excited oscillations allows to compute the average power \cite{Ventura19} needed to maintain the real sliding-mode. Numerical simulations are presented to confirm the results.

\section{Preliminaries}\label{Sec3}
\subsection{Describing Function approach}
The Describing Function (DF) approach allows to find a possible periodic motion caused by the actuator dynamics  (\ref{Actuator}) into the system (\ref{W_Actuator}) with control law (\ref{Control}). Taking into account just the first-harmonic of the steady state response,
\begin{equation}\label{Limit_Cycle}
\begin{array}{rcl}
x_1(t) & \approx & A \sin(\omega t)\,, \\
x_2(t) & \approx & A\omega \cos(\omega t)\,,
\end{array}
\end{equation}
where $A$ is the amplitude and $\omega$ the frequency. The expression (\ref{Limit_Cycle}) approximates accurately of the steady state behavior, due to linear system (\ref{W_Actuator}) has sharp low-pass filter characteristics (proper transfer function) \cite{Gelb68,Atherton75}, i.e. $|W(j\omega)|>>|W(jn\omega)|$, for $n=2, \,3, ...$, where $\omega$ is the frequency of self-excited oscillations. \\

Parameters of the periodic motion (\ref{Limit_Cycle}) can be found as an intersection point of the Nyquist plot $W(j\omega)$ and the negative reciprocal Describing Function $N(A,\omega)$ of the controller (\ref{Control}). This means that chattering parameters can be predicted by solving the Harmonic Balance (HB) equation \cite{Gelb68,Atherton75},
\begin{equation}\label{HBE}
N(A,\omega)W(j\omega) = -1 \,.
\end{equation}  
The DF of the non-linearity $u(t)$ is defined as the first harmonic Fourier series of the periodic control (\ref{Control}) divided by the amplitude $A$ of the oscillatory state (\ref{Limit_Cycle}), i.e.
\begin{equation}\label{DF_def}
N(A,\omega) = \dfrac{a_1+jb_1}{A} \,,
\end{equation}
where the Fourier coefficients are
\begin{eqnarray}
a_1 & = & \frac{\omega}{\pi}\displaystyle\int_{0}^{\frac{2\pi}{\omega}} \hspace{-2mm} u(t) \sin (\omega t) \text{d} t \,, \label{DF_coef_a1} \\
b_1 & = & \frac{\omega}{\pi}\displaystyle\int_{0}^{\frac{2\pi}{\omega}} \hspace{-2mm} u(t) \cos (\omega t) \text{d} t \,. \label{DF_coef_b1}
\end{eqnarray}

\subsection{Average Power}
The estimation of the chattering parameters, amplitude (\ref{Amp}) and frequency (\ref{Freq}) of the periodic motion (\ref{Limit_Cycle}), allows to compute the instantaneous power \cite{Khalil02} 
\begin{equation}\label{P_inst}
p(t) = \bar{u}(t)x_1(t) = \frac{A^2\omega}{2} \sin(2\omega t) \,. 
\end{equation}
Let assume, for example, that system (\ref{System}) models an electrical circuit where $x_1$ is the current and $\bar{u}$ is the voltage. Then, the average power \cite{Ventura18,Ventura19} needed to maintain the \textit{real} second order sliding-mode can be computed, for each period $T = \frac{2\pi}{\omega}$ predicted by HB, as 
\begin{equation}\label{P_avg}
P = \frac{1}{T} \int_0^{T} |p(t)| \text{d}t = \dfrac{4A^2\omega}{\pi} \,. 
\end{equation}
It should be mentioned that the average power (\ref{P_avg}) can only be computed taking into account the presence of the actuator dynamics, because there is no chattering in \textit{ideal} sliding mode by definition. 

\begin{figure}[t]
	\begin{center}
		\vspace{0mm}
		\includegraphics[scale=0.25]{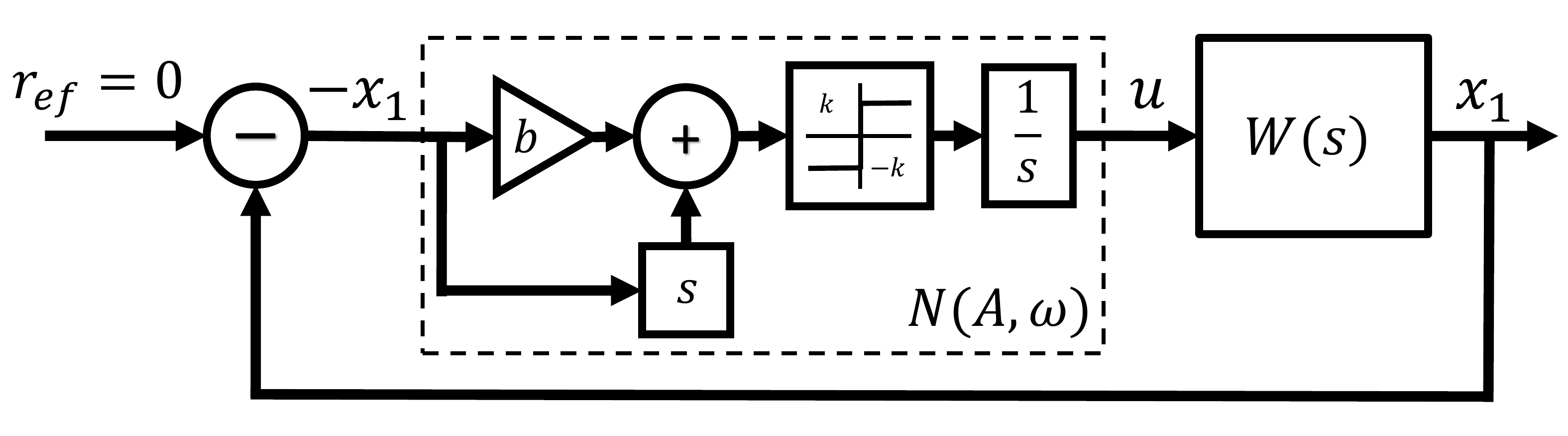}   
	\end{center}
	\vspace{-2mm}
	\caption{Describing Function $N(A,\omega)$ of the LSV-LCSMC (\ref{Sigma1}), (\ref{Control}), with oscillating steady-state response (\ref{Limit_Cycle}).}\label{Bloques_DF}
	\vspace{0mm}
\end{figure}

\begin{figure}[t]
	\centering
	\subfigure[LSV-LCSMC]{\includegraphics[width=70mm]{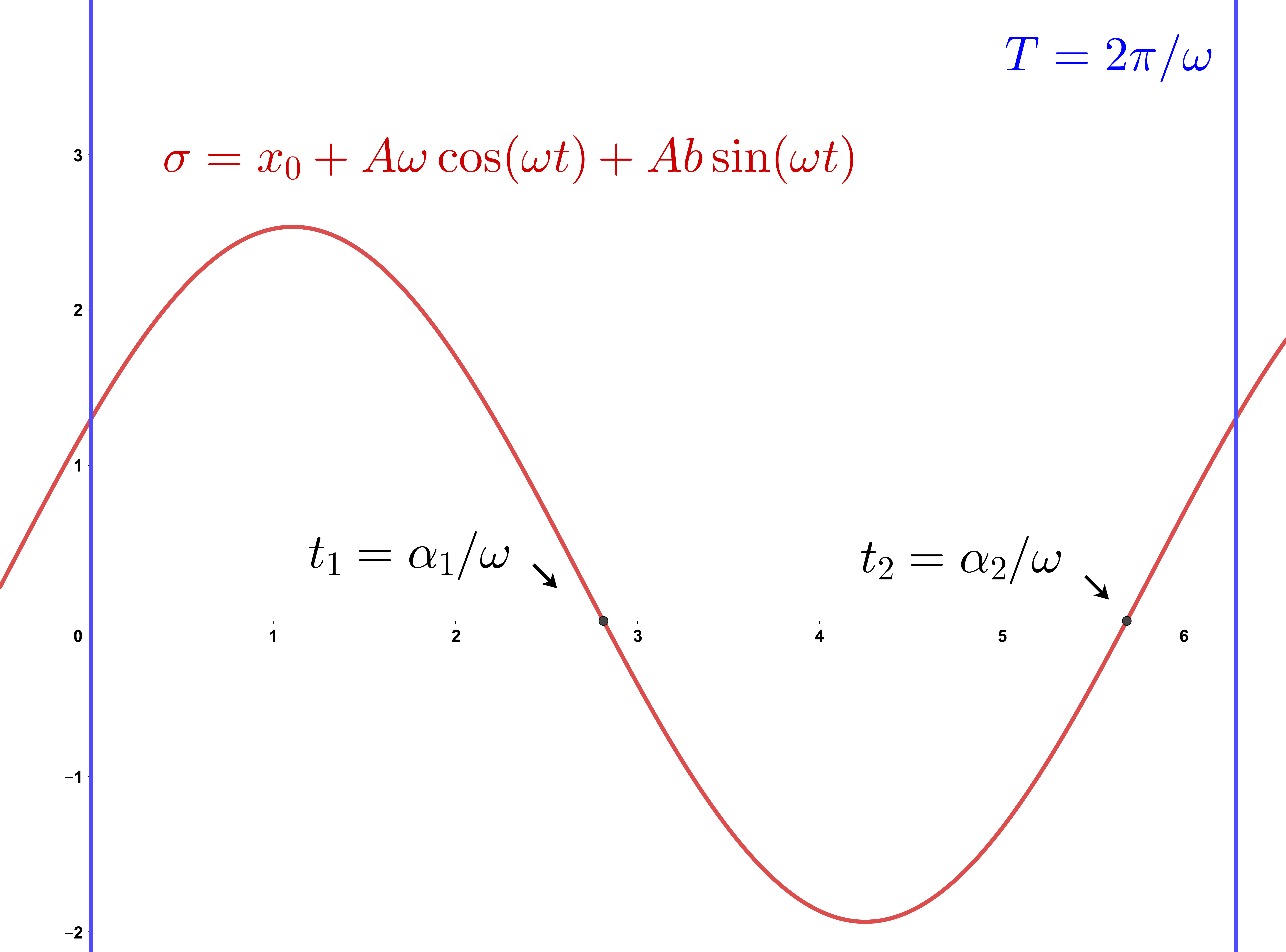}}
	\subfigure[TSV-LCSMC]{\includegraphics[width=70mm]{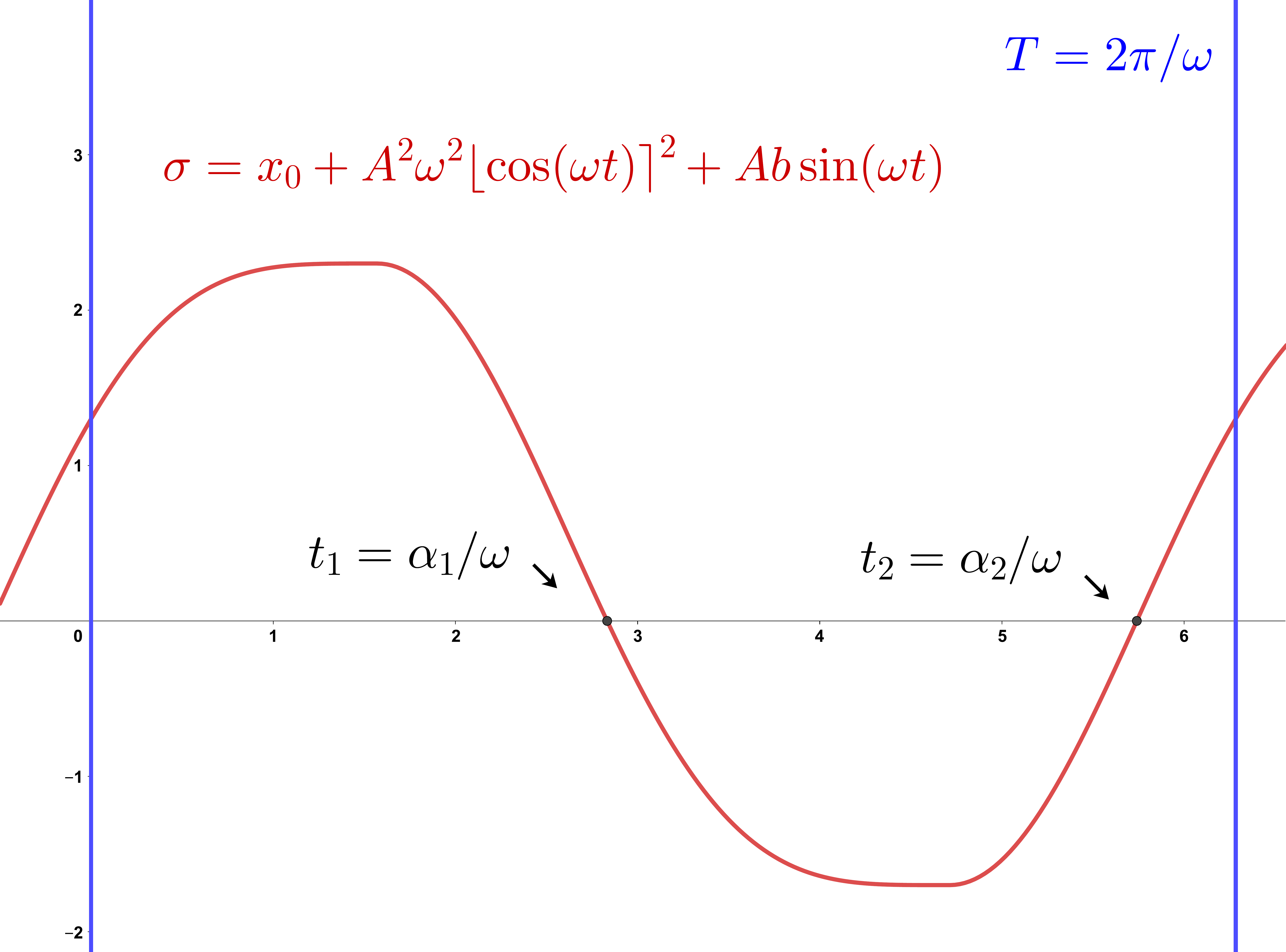}}
	\caption{Sign changes on the sliding variable for LSV-LCSMC and TSV-LCSMC in each period of time.}
	\label{fig:cambiossigno}
\end{figure}

\section{Calculation of the Describing Functions}\label{Sec4}
\subsection{Describing Function of the LSV-LCSMC}
Following Figure \ref{Bloques_DF} where the LCSMC  (\ref{Control}) with LSV (\ref{Sigma1}) is used to drive the system (\ref{W_Actuator}). Then, the steady-state behavior of the sliding variable, taking into account (\ref{Limit_Cycle}), is such that
\begin{equation*}
\sigma = x_2+b\,x_1=A\omega \cos (\omega t)+Ab\sin (\omega t) \,.
\end{equation*}
In order to compute the Fourier coefficients in (\ref{DF_def}), since the control law is defined by a sign function, only the points where the surface change it sign are required, i.e.
\begin{equation*}
\sigma = \omega \cos (\omega t_{1,2})+b\sin (\omega t_{1,2})=0 \,.
\end{equation*}
There are only two sign changes in the sliding variable (see Figure \ref{fig:cambiossigno}(a)) for each time period and they can be found as
\begin{equation*}
\begin{split}
\sin (\omega t_{1,2})= \pm \dfrac{\omega}{\sqrt{\omega^2+b^2}} \,, \\
\cos (\omega t_{1,2})= \pm \dfrac{b}{\sqrt{\omega^2+b^2}} \,.
\end{split}
\end{equation*}
Note that at the first sign change ($t_1=\alpha_1/\omega$), $\sin (\omega t_1)>0$ and $\cos (\omega t_1)<0$, and then
at the second change ($t_2=\alpha_2/\omega$), $\sin (\omega t_2)<0$ and $\cos(\omega t_2)>0$. Then, coefficient (\ref{DF_coef_a1}) can be computed as
\begin{equation*}
\begin{split}
a_1&=\dfrac{\omega}{\pi}\int_{0}^{\frac{2\pi}{\omega}}\brackc{\omega \cos (\omega t)+b\sin (\omega t)}{0}\sin (\omega t)\text{d}t \, \\
&=\dfrac{\omega}{\pi} \left(\int_{0}^{\frac{\alpha_1}{\omega}}\hspace{-1.7mm}\sin (\omega t)\text{d}t-\int_{\frac{\alpha_1}{\omega}}^{\frac{\alpha_2}{\omega}}\hspace{-1.5mm}\sin (\omega t)\text{d}t+\int_{\frac{\alpha_2}{\omega}}^{\frac{2\pi}{\omega}}\hspace{-1.5mm}\sin (\omega t)\text{d}t\right)\hspace{-1mm}\, \\
&=\dfrac{1}{\pi}(2\cos (\alpha_2)-2\cos (\alpha_1)) =\dfrac{4\,b}{\pi\sqrt{\omega^2+b^2}} \,,
\end{split}
\end{equation*}
On the other hand, coefficient (\ref{DF_coef_b1}) can be calculated as
\begin{equation*}
\begin{split}
b_1&=\dfrac{\omega}{\pi}\int_{0}^{\frac{2\pi}{\omega}}\brackc{\omega \cos (\omega t)+b\sin (\omega t)}{0}\cos (\omega t)\text{d}t \, \\
&=\dfrac{\omega}{\pi} \left(\int_{0}^{\frac{\alpha_1}{\omega}}\hspace{-1.7mm}\cos (\omega t)\text{d}t-\int_{\frac{\alpha_1}{\omega}}^{\frac{\alpha_2}{\omega}}\hspace{-1.5mm}\cos (\omega t)\text{d}t+\int_{\frac{\alpha_2}{\omega}}^{\frac{2\pi}{\omega}}\hspace{-1.5mm}\cos (\omega t)\text{d}t\right) \hspace{-1mm}\, \\
&=\dfrac{1}{\pi}(2\sin (\alpha_1)-2\sin (\alpha_2))=\dfrac{4\,\omega}{\pi\sqrt{\omega^2+b^2}} \,.
\end{split}
\end{equation*}
Since the LCSMC (\ref{Control}) is defined as an integral controller, the integrator transfer function must be added to the DF, and it ends as
\begin{equation}\label{DF1}
\begin{split}
N(A,\omega)&=\dfrac{4k}{\pi A\sqrt{\omega^2+b^2}}\left(b+j\omega\right)\left(\dfrac{1}{j\omega}\right) \, \\
&=\dfrac{4k}{\pi A\sqrt{\omega^2+b^2}}\left(1-j\dfrac{b}{\omega}\right) \,.
\end{split}
\end{equation}
\subsection{Describing Function of the TSV-LCSMC}
Now, if the terminal sliding variable (\ref{Sigma2}) is chosen, the virtual output will have a behavior as
\begin{equation*}
\sigma =A^2\omega^2 \brackc{\cos (\omega t)}{2}+b A\sin (\omega t) \,.
\end{equation*}
The TSV (\ref{Sigma2}) is a nonlinear surface, that is why from the Figure \ref{fig:cambiossigno}(b), $\sigma$ does not have a sinusoidal form. 

Following the same steps made for the linear variable, the DF for the TSV-LCSMC can be computed as
\begin{equation}\label{DF2}
\begin{split}
N(A,\omega)=&\dfrac{2k}{\pi A^2\omega^3}\left(\sqrt{b^2+4A^2\omega^4}-b\right.\\ &\left.-j\sqrt{2b\left(\sqrt{b^2+4A^2\omega^4}-b\right)} \right).
\end{split}
\end{equation}

\begin{figure}[t]
	\begin{center}
		\vspace{0mm}
		\includegraphics[scale=0.55]{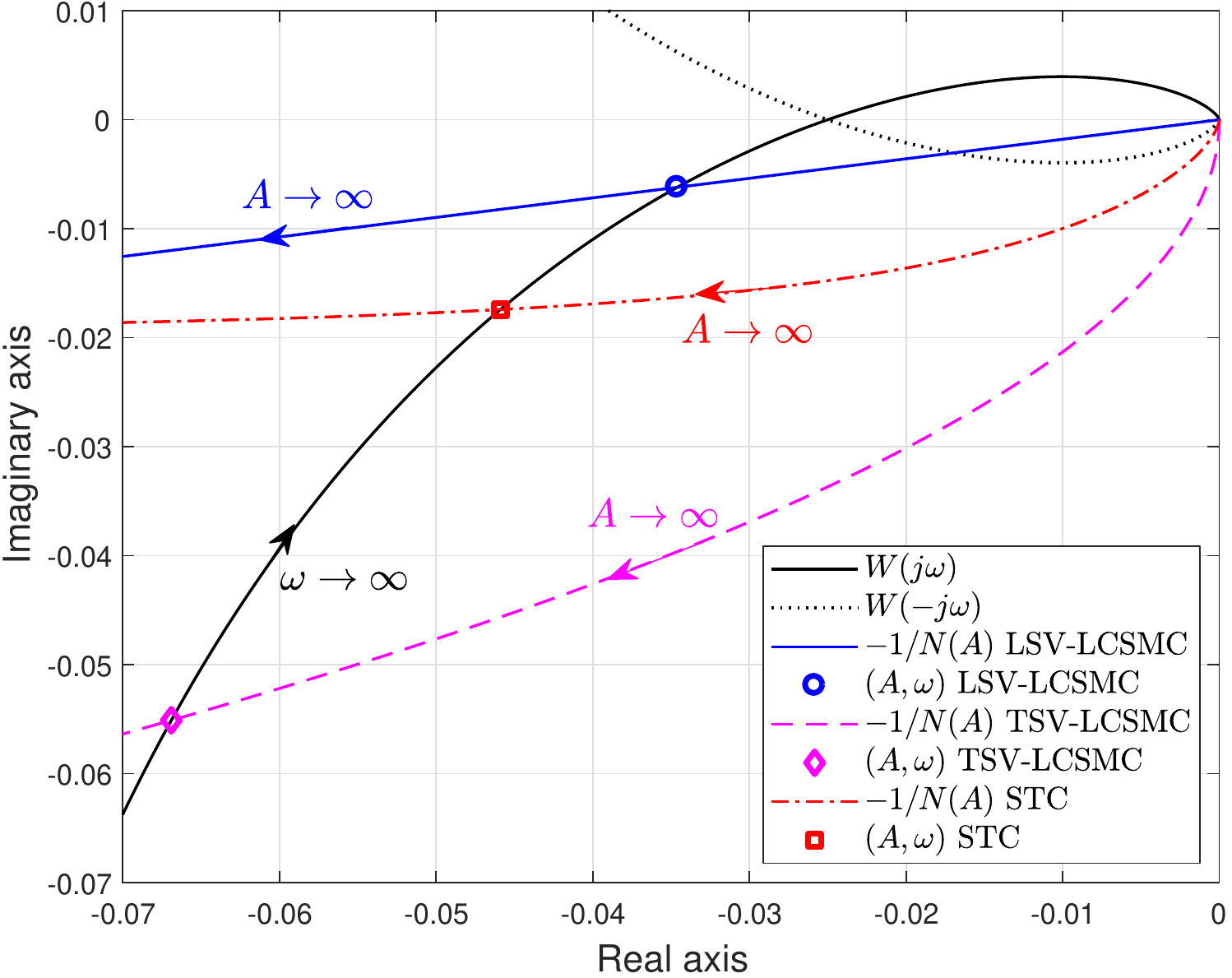}   
	\end{center}
	\vspace{-2mm}
	\caption{Graphical solution of the HB (\ref{HBE}) equation for LSV-LCSMC, TSV-LCSMC and STC with parameters $k=k_2=1.1\Delta$, $k_1=2\Delta^{1/2}$, $\Delta=5$, $b=3$ and $\mu=0.05$.}\label{HBGraph}
	\vspace{0mm}
\end{figure}

\begin{figure}[t]
	\begin{center}
		\vspace{0mm}
		\includegraphics[scale=0.4]{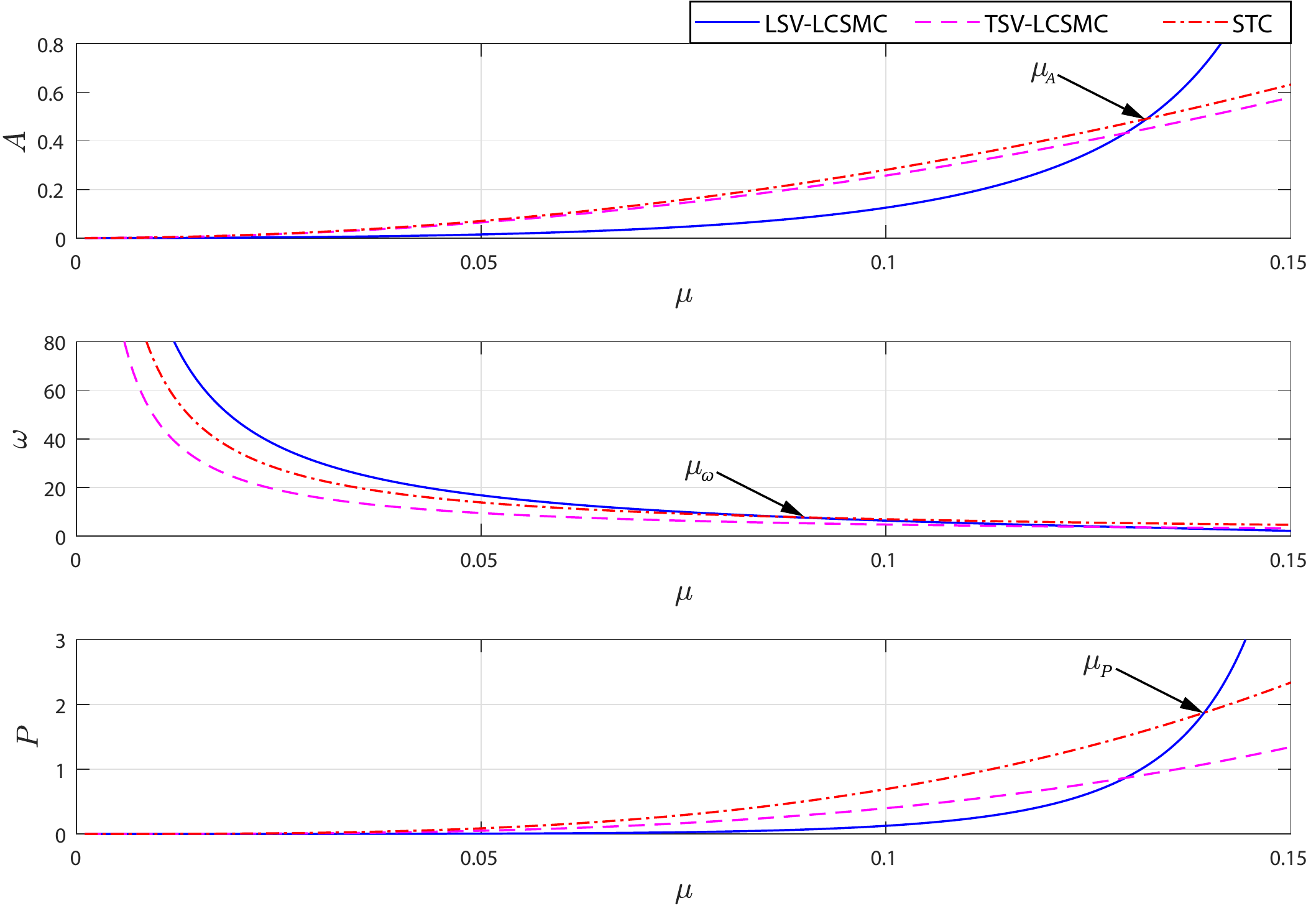}   
	\end{center}
	\vspace{-2mm}
	\caption{Predicted chattering parameters: amplitude, frequency and average power generated by the LSV-LCSMC, the TSV-LCSMC and STC, with $k=k_2=1.1\Delta$, $k_1=2\Delta^{1/2}$, $\Delta=5$ and $b=3$, for $\mu\in\left[0,\frac{1}{2b}\right)$.}\label{Chatt_teo}
	\vspace{0mm}
\end{figure}

\section{Chattering predictions for the Lipschitz continuous sliding-mode controllers}\label{Sec5}
A critically damped model of the actuator dynamics in (\ref{ActuatorModel}) is usually considered \cite{Utkin16,Ventura19,Boiko07,ByF05}
\begin{equation}\label{Actuator}
G_a (s) = \frac{1}{(\mu s + 1)^2} \,, 
\end{equation} 
where $s$ is the complex variable of Laplace transformation, and the small parameter $\mu>0$. The linear system conformed by the cascade connection of the critically damped actuator (\ref{Actuator}) and the plant (\ref{System}) (see Figure \ref{Bloques}) can be written as
\begin{equation}\label{W_Actuator}
W(s) = \dfrac{1}{s(\mu s + 1)^2} \,,
\end{equation}
where $f(t) = 0$ is assumed. \\
\subsection{Analysis of fast-motions in dynamically perturbed systems driven by LCSMC}
\begin{itemize}
	\item \textbf{LSV-LCSMC} 
\end{itemize}
The HB equation (\ref{HBE}) for the system (\ref{W_Actuator}) and the DF (\ref{DF1}) of the LSV-LCSMC can be rewritten as 
\begin{equation}\label{HBE1}
\dfrac{4k}{\pi A\sqrt{\omega^2+b^2}} - j \dfrac{4kb}{\pi A\omega\sqrt{\omega^2+b^2}} = 2 \, \mu \, \omega^2  + j \, \omega \, (\mu^2 \, \omega^2 - 1) \,, \vspace{-2mm}
\end{equation}
whose solutions are 
\begin{align}
A & = \mu^2 \left(\dfrac{2k}{\pi(1-2\mu b)(1-\mu b)} \right) \,, \label{Amp} \\
\omega & = \, \dfrac{1}{\mu}\left(\sqrt{1-2\mu b}\right) \,. \label{Freq}  
\end{align}
\begin{remark}\label{rem1}
	\textit{It should be noted that parameter $b$ can not be chosen arbitrarily, if $b \ge\frac{1}{2\mu}$ the system can lose stability in the presence of fast actuator. This fact is reflected in equation (\ref{Amp}), where the amplitude tends to infinity for $b\rightarrow\frac{1}{2\mu}$. It is reasonable because for $b \ge\frac{1}{2\mu}$ the sliding surface becomes faster than the actuator.}
\end{remark}

The graphical solution of the HB equation (\ref{HBE1}) is also plotted in Figure \ref{HBGraph}. The average power (\ref{P_avg}) can be computed by replacing the estimated parameters (\ref{Amp}) and (\ref{Freq}) and has the form
\begin{equation}\label{Pot}
P = \mu^3\left(\dfrac{16k^2}{\pi^3(1-2\mu b)^{3/2}(1-\mu b)^2}\right) \,.
\end{equation}
Figure \ref{Chatt_teo} shows the predicted chattering parameters: amplitude (\ref{Amp}), frequency (\ref{Freq}) and average power (\ref{Pot}) generated by the LSV-LCSMC (\ref{Sigma1}), (\ref{Control}), in the system (\ref{W_Actuator}) for several values of the actuator time constant $\mu>0$. It could be note that as the actuator dynamics be slower ($\mu$ grows), the amplitude of chattering and the average power both increase. Additionally, there exist an actuator time-constant for which the closed-loop trajectories are unstable (see Remark \ref{rem1}), this value is $\mu = 1/6$ for the gains $k=1.1\Delta$, $\Delta=5$ and $b=3$.

\begin{itemize}
	\item \textbf{TSV-LCSMC} 
\end{itemize}
The HB equation (\ref{HBE}) for the system (\ref{W_Actuator}) and the DF (\ref{DF2}) of the TSV-LCSMC can only be solved numerically. One can see such solution from the Figure \ref{HBGraph}. Figure \ref{Chatt_teo} shows the predicted chattering parameters generated by the TSV-LCSMC (\ref{Sigma2}), (\ref{Control}), in the system (\ref{W_Actuator}) for several values of the actuator time-constant $\mu>0$.

\subsection{Comparison with Super-Twisting controller}
In this paper, a brief comparison of LSV-LCSMC (\ref{Sigma1}), (\ref{Control}), and TSV-LCSMC (\ref{Sigma2}), (\ref{Control}), is done. Also a comparison with  by the STC \cite{Levant98} having the form
\begin{equation}\label{STA}
u  =  -k_1 \lceil x_1 \rfloor^{1/2} + v \,, \quad
\dot{v}  =  -k_2 \lceil x_1 \rfloor^0 \,. 
\end{equation}
is performed. 
For perturbed systems of relative degree one, the STC parameters can be chosen $k_1>\sqrt{8(k_2+\Delta)}$, $k_2>\Delta$ ensuring finite-time convergence of the output and its derivative to zero \cite{Seeber18}.

\subsubsection{Comparison of fast-motions}
The STC produces a continuous but not Lipschitz control signal (with infinite slope at the origin), then, it also causes chattering in the presence of fast-parasitic dynamics \cite{ByF05}. The STC has two sources of chattering related to their non-linearities. The DF for the STC (\ref{STA}) has the form \cite{ByF05}
\begin{equation}\label{DF_STA}
N(A,\omega) = \dfrac{2\alpha_1 k_1}{\pi A^{1/2}} + \dfrac{1}{j\omega}\left( \dfrac{4 k_2}{\pi A} \right) \,,
\end{equation}
where $\alpha_1 \approx 1.748$. The chattering parameters predicted by HB of the system \ref{W_Actuator} in closed-loop with the STC (\ref{STA})  are \cite{Ventura18,Ventura19},
\begin{equation}\label{STA_chatt}
\begin{split}
A =& \mu^2 \, \left(\dfrac{\alpha_1^2 k_1^2 + 4\pi k_2}{\pi\alpha_1k_1} \right)^2 \,, \\
\omega =& \frac{1}{\mu} \, \left( \dfrac{\alpha_1^2 k_1^2}{\alpha_1^2 k_1^2 + 4\pi k_2} \right)^{1/2} \,, \\ 
P = & \mu^3 \, \left( \dfrac{4\left( \alpha_1^2 k_1^2 +  4\pi k_2 \right)^{7/2} }{\pi^5\alpha_1^3 k_1^3} \right) \,. \vspace{0mm}
\end{split}
\end{equation}
Following \cite{Ventura19}, the gains $k_1 \approx 2\Delta^{1/2}$ and $k_2 = 1.1\Delta$ can be chosen in order to minimize the amplitude of oscillations and the average power in (\ref{STA_chatt}). The Figure \ref{Chatt_teo} shows the chattering parameters (\ref{STA_chatt}) of the steady state response for system (\ref{W_Actuator}), governed by STC (\ref{STA}), for $\mu\in\left[0,\frac{1}{2b}\right)$, $\frac{1}{2b}=\frac{1}{6}$. 

\begin{remark}
	\textit{Comparing the predicted amplitudes for LSV-LCSMC (\ref{Amp}) and for STC in (\ref{STA_chatt}), the value of $\mu$ for which the amplitude of oscillations caused by LSV-LCSMC and by STC are the same can be found as
		\begin{equation} \label{Mu_IRC_STA_A}
		\mu_{A_{1,2}} = \dfrac{3\pm\sqrt{9-8\gamma}}{4b} \,,
		\end{equation}
		where $\gamma=\dfrac{( (\alpha_1k_1)^2+4\pi k_2 )^2 - 2\pi(\alpha_1k_1)^2k}{( (\alpha_1k_1)^2+4\pi k_2 )^2}$.}\vspace{1mm}
\end{remark}
For $k_1=2\Delta^{1/2}$, $k_2=k=1.1\Delta$, $\Delta=5$ and $b=3$, expression (\ref{Mu_IRC_STA_A}) becomes
\begin{eqnarray}
\mu_{A_{1}} = & 0.1323 \,, \label{Mu_A_1}\\
\mu_{A_{2}} = & 0.3677 \,, \label{Mu_A_2}
\end{eqnarray}
the actuator time constant (\ref{Mu_A_2}) does not satisfy the stability condition $2b\mu<1$ imposed from the parameter selection $b=3$ (see Remark \ref{rem1}). The value (\ref{Mu_A_1}) is shown in the first plot of Figure \ref{Chatt_teo}.

\begin{remark}
	\textit{Comparing the predicted frequencies for LSV-LCSMC (\ref{Freq}) and for STC in (\ref{STA_chatt}), the value of $\mu$ for which frequency of oscillations caused by LSV-LCSMC and  STC are same can be calculated as
		\begin{equation} \label{Mu_w}
		\mu_\omega = \dfrac{4\pi k_2}{2b((\alpha_1k_1)^2+4\pi k_2)} \,.
		\end{equation}}\vspace{-2mm}
\end{remark}
So $\mu_\omega = 0.0885$ with the gains $k_1=2\Delta^{1/2}$, $k_2=k=1.1\Delta$, $\Delta=5$ and $b=3$. The value of parameter (\ref{Mu_w}) is marked in the second plot of Figure \ref{Chatt_teo}.

\begin{figure}[t]
	\begin{center}
		\vspace{0mm}
		\includegraphics[scale=0.4]{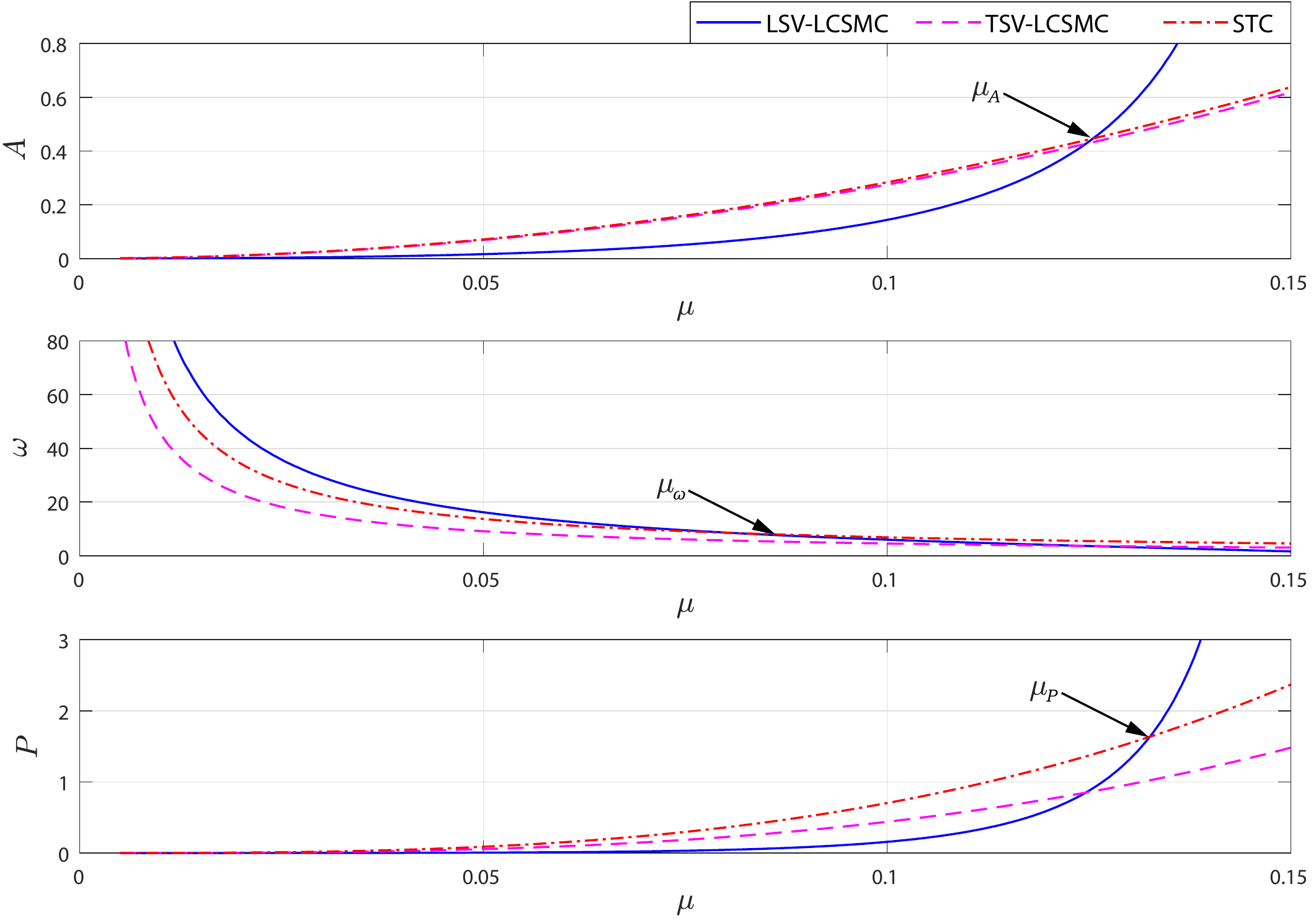}   
	\end{center}
	\vspace{-2mm}
	\caption {Numerically calculated chattering parameters: amplitude, frequency and average power generated by the LSV-LCSMC, the TSV-LCSMC and STC, with $k=k_2=1.1\Delta$, $k_1=2\Delta^{1/2}$, $\Delta=5$ and $b=3$, for $\mu\in\left[0,\frac{1}{6}\right)$.}\label{Chatt_sim}
	\vspace{0mm}
\end{figure} 

\begin{remark}
	\textit{Comparing the predicted average power for LSV-LCSMC (\ref{Pot}) and for STC in (\ref{STA_chatt}), the value of $\mu$ for which the average power required to maintain the real sliding mode with LSV-LCSMC is the same than one required using STA can be found solving the equation}
	\begin{equation} \label{Mu_IRC_STA_P}
	(1-2\mu b)^3(1-\mu b)^4 = \frac{16\pi^4k^4(\alpha_1k_1)^6}{((\alpha_1k_1)^2+4\pi k_2)^7} \,.
	\end{equation}
\end{remark}
For $k_1=2\Delta^{1/2}$, $k_2=k=1.1\Delta$, $\Delta=5$ and $b=3$, then solutions of (\ref{Mu_IRC_STA_P}) are
\begin{eqnarray}
\mu_{P_{1}} = & 0.1392 \,, \label{Mu_P_1}\\
\mu_{P_{2,3}} = & 0.1750\pm j0.0339 \,, \label{Mu_P_23}\\
\mu_{P_{4,5}} = & 0.3050\pm j0.0475 \,, \label{Mu_P_45}\\
\mu_{P_{6,7}} = & 0.3669\pm j0.0273 \,. \label{Mu_P_67}
\end{eqnarray}
The actuator time constant (\ref{Mu_P_1}) is the unique solution such that the average power of oscillations caused by the LSV-LCSMC is equal those one caused by the STC. This value is also marked in the third plot of Figure \ref{Chatt_teo}.

\begin{figure}[t]
	\begin{center}
		\vspace{0mm}
		\includegraphics[scale=0.4]{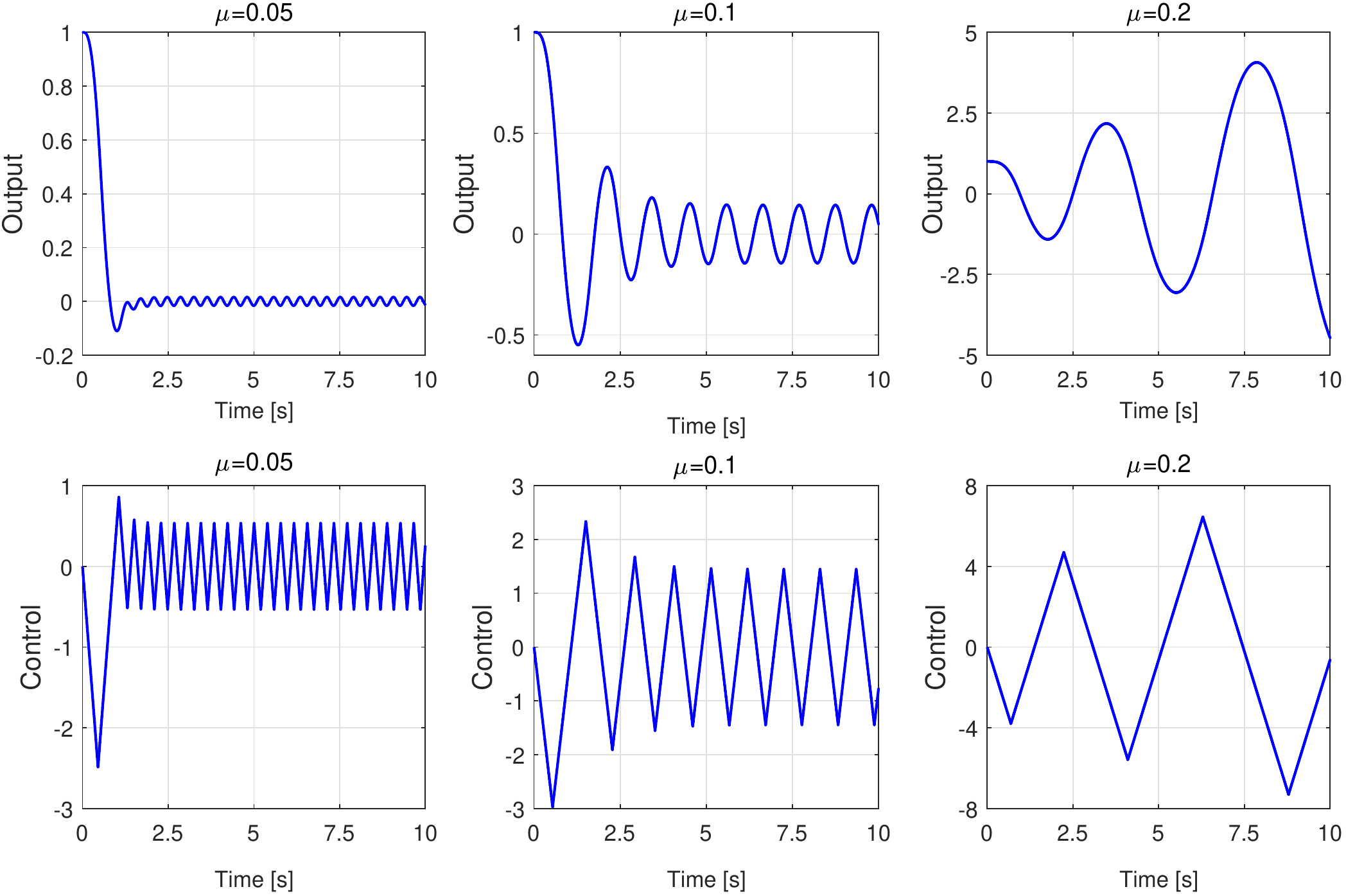}   
	\end{center}
	\vspace{-4mm}
	\caption{Closed-loop behavior of the system (\ref{W_Actuator}) in with the LSV-LCSMC (\ref{Sigma1}), (\ref{Control}), fixing the parameters $b=3$, $k=1.1\Delta$, $\Delta=5$ and for three different value of $\mu$. For $\mu=0.2$ the system becomes unstable.}\label{Simu1}
	\vspace{0mm}
\end{figure}

\section{Numerical Testing}\label{Sec6}
Consider the following simulation framework: the Euler's integration method with constant step $\tau=10^{-4}$ is used to solve the closed-loop (\ref{CL}), the fixed value of time-constant $\mu = 0.05$ is considered for the actuator dynamics (\ref{W_Actuator}). The upperbound $\Delta=5$ is taken for simplicity.

\subsection{Fast-Motions}
The unperturbed response is conformed only by the oscillations caused by the actuator dynamics in (\ref{W_Actuator}). Then, the perturbation $f(t)=0$ is fixed for simulations. Let the gains $k_1=2\Delta^{1/2}$, $k_2=k=1.1\Delta$, $\Delta=5$ and $b=3$. As mentioned in Remark \ref{rem1}, for $\mu=\frac{1}{2b}$ the system with LSV-LCSMC become unstable for any selected $b>0$. This can be observed in the Figure \ref{Simu1} where $b=3$ was selected. On the left plot, the output of LSV-LCSMC oscillates close to zero,since the actuator is fast enough ($2b\mu=0.3<1$). For $\mu=0.1$ (middle plot) $2b\mu=0.6 <1$ the system trajectories still remains in a vicinity of zero, but the oscillations of the output become bigger. Finally, for $2b\mu = 1.2>1$ (right plot) the trajectories of the system diverge.

Figure \ref{Chatt_sim} shows the chattering parameters measured from simulations for the dynamically perturbed system (\ref{W_Actuator}) in closed-loop with the LSV-LCSMC, TSV-LCSMC and STC, respectively, and for several values of the actuator time-constant $\mu>0$. The values $\mu_A$, $\mu_\omega$, $\mu_P$ for which the amplitude, frequency and average power generated by the STC and LSV-LCSMC are the same can be obtained from simulations,
\begin{align}
\mu_A = 0.1255 \,, \\
\mu_\omega = 0.0811 \,, \\
\mu_P = 0.1325 \,,
\end{align}
which are well predicted by the HB methodology, if they are compared against the theoretical predictions in (\ref{Mu_A_1}), (\ref{Mu_w}) and (\ref{Mu_P_1}), respectively.

\begin{figure}[t]
	\begin{center}
		\vspace{0mm}
		\includegraphics[scale=0.4]{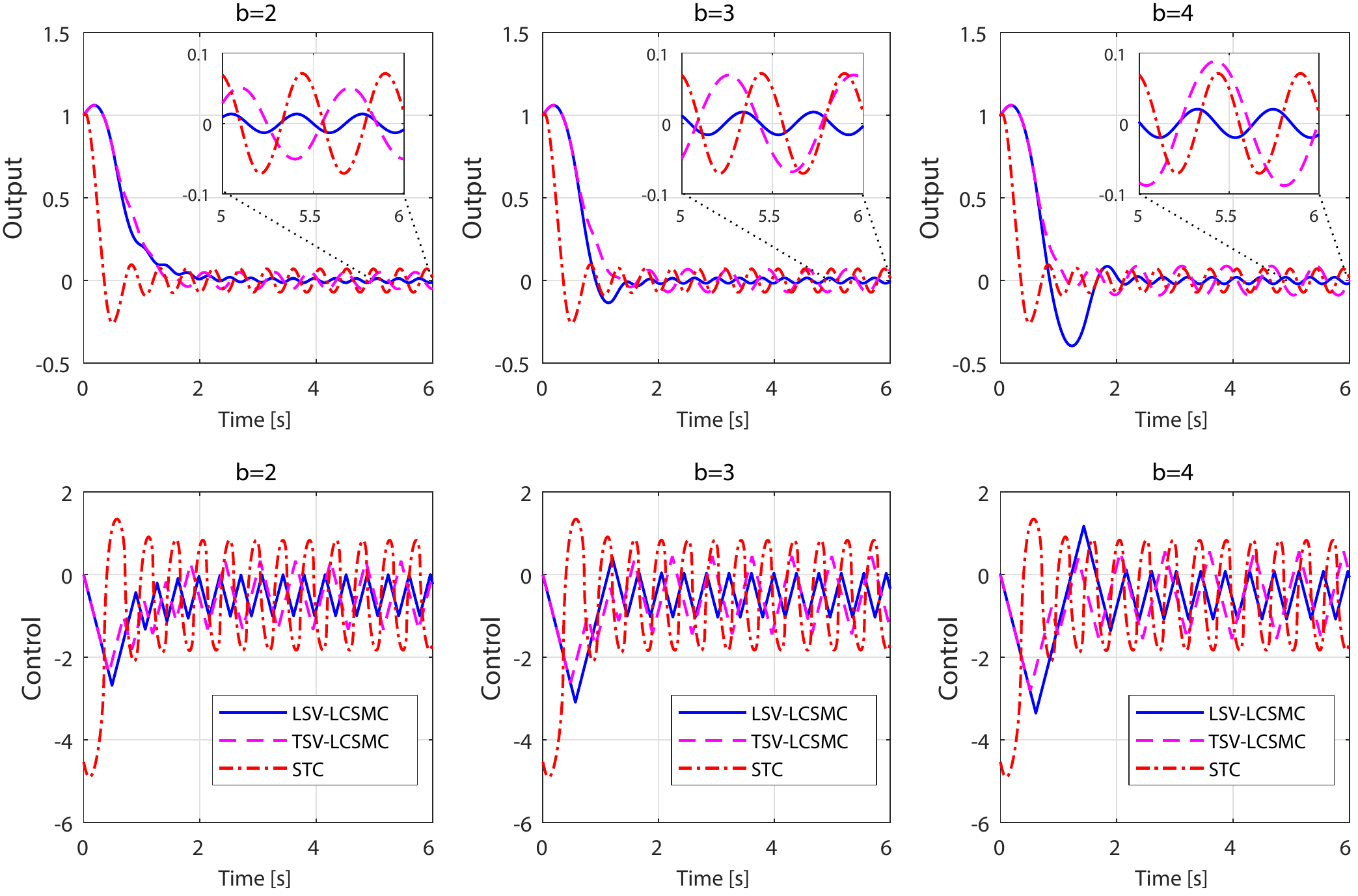}   
	\end{center}
	\vspace{-4mm}
	\caption{Closed-loop behavior of the system (\ref{W_Actuator}) with the LSV-LCSMC and the TSV-LCSMC, in comparison with using the STC, for $b=2,3,4$, $k=k_2=1.1\Delta$, $k_1=2\Delta^{1/2}$, $\Delta=5$ and $\mu=0.05$.}\label{Simu2}
	\vspace{0mm}
\end{figure}

In Figure \ref{Simu2}, it is shown the behavior of the system (\ref{W_Actuator}) in closed-loop with the three aforementioned controllers. Considering the optimal gains for STC and different values of the constant $b$ for the LCSMC. In the left plot could be observed the oscillations produced for $b=2$, both LCSMC produce chattering with smaller amplitude than the STC, but with a bigger convergence time. Choosing $b=3$ (middle plot), the transient time for both LCSMC is reduced but the oscillations caused by TSV-LCSMC has almost the same size that one produced by the STC. Finally, if  $b=4$ is selected (right plot), the oscillations produced by TSV-LCSMC are larger than one caused by the STC. On the other hand, the oscillations produced by LSV-LCSMC remains smaller than one caused by the STC, but the overshoot is increased (it can even become unstable) as the parameter $b$ grows.

\section{Conclusions}\label{Sec7}


In this work, the Describing Functions for the LSV-LCSMC and TSV-LCSMC are calculated for the first time. These calculations are used to predict the chattering parameters: amplitude, frequency and the average power needed to maintain the system into real sliding-modes, using the Harmonic Balance approach. These predictions allow to analyze and compare the chattering parameters produced by LSV-LCSMC and TSV-LCSMC, between them, and with the STC.

Performed comparisons confirm the hypothesis formulated by Prof. Utkin, that the finite-time convergence is not a critical issue for LCSMC. 
The amplitude of chattering generated by LSV-LCSMC is smaller than one caused by TSV-LCSMC, at least for fast-actuators. 
However, it is necessary to remark if the sliding dynamics of the LSV-LCSMC closed-loop is of similar speed as the actuators dynamics, the system can loose even practical stability.

\section*{Acknowledgment}
The authors are grateful for the financial support of CONACyT (Consejo Nacional de Ciencia y Tecnolog\'{i}a): CVU 631139, 631266, 711867; Grant 282013; PAPIIT-UNAM (Programa de Apoyo a Proyectos de Investigaci\'{o}n e Innovaci\'{o}n Tecnol\'{o}gica) IN 115419.

\end{document}